\documentclass[preprint,showpacs,preprintnumbers,amsmath,amssymb,nofootinbib]{revtex4}
\usepackage{epsfig}
\DeclareFontFamily{OT1}{rsfs}{}
\DeclareFontShape{OT1}{rsfs}{m}{n}{<5> rsfs5 <7> rsfs7 <10> rsfs10}{}
\DeclareSymbolFont{mathrsfs}{OT1}{rsfs}{m}{n}
\DeclareSymbolFontAlphabet{\mathrsfs}{mathrsfs}
\newcommand{\A}{\mathbf{A}}
\newcommand{\K}{\mathbf{k}}

\newcommand{\kk}{{k_{\parallel}}}
\renewcommand{\k}{{k_{\perp}}}

\newcommand{\h}{\hat{H}}

\newcommand{\arccosh}{ \mathrm{arccosh}\,}
\newcommand{\arcsinh}{ \mathrm{arcsinh}\,}
\renewcommand{\Im}{\mathrm{Im}\,}

\begin{document}
\title{Instantaneous Multiphoton Ionization Rate and Initial Distribution of Electron Momenta}

\author{Denys I. Bondar}
\address{Department of Physics and Astronomy, University of Waterloo,
200 University Avenue West, Waterloo, Ontario N2L 3G1, Canada}
\address{National Research Council of Canada, 100 Sussex Drive, Ottawa, Ontario
K1A 0R6, Canada}
\email{dbondar@sciborg.uwaterloo.ca}

\date{\today}

\begin{abstract}
The Yudin-Ivanov formula [Phys. Rev. A {\bf 64}, 013409 (2001)]
is generalized such that the most general analytical expression for single-electron spectra, which includes the dependence on the instantaneous laser phase, is obtained within the strong field approximation. No assumptions on the momentum of the electron are made.  Previously known formulas for single-electron spectra can be obtained as approximations to the general formula. \\
\\
{\it This preprint is published in Phys. Rev. A {\bf 78}, 015405 (2008).}
\end{abstract}


\maketitle

Although an initial theoretical understanding of strong field ionization was put forth by Keldysh \cite{Keldysh_1965} as early as 1964, many questions remain unresolved.  As far as single-electron ionization in the presence of a linearly polarized laser field is concerned, there are two important topics.  The first, namely the ionization rate as a function of instantaneous laser phase, was studied in depth by Yudin and Ivanov \cite{Yudin_2001} (see also \cite{Uiberacker_2007, Kienberger_2007}).  However, their result assumes zero initial momentum of the liberated electron. Effects due to nonzero initial momentum have yet to be included.  The second topic pertains to the single-electron spectra, that is, the ionization rate as a function of the final momentum of an electron.  Despite much work on this topic (see, for example, Refs. \cite{Krainov_2003, Goreslavskii_2005, Popov_2004} and references therein) a universally accepted formula is lacking and the discussion is still ongoing.  Among the most accurate results, Goreslavskii {\it et al.} \cite{Goreslavskii_2005} have obtained an expression for the complete single-electron ionization spectrum, but without consideration of laser phase.  The present paper derives a more general formula [see Eq. (\ref{GeneralRate})] that includes the dependences on both the instantaneous laser phase and the final electron momentum.

It is very convenient to formulate the Keldysh theory in terms of the Dykhne adiabatic approximation, as was presented for in \cite{Delone_1985}.  According to the Dykhne method \cite{Dykhne_1962, Davis_1976, Chaplik_1964} (see also \cite{Landau_1977, Delone_1985}), if the Hamiltonian of a system $\h(t)$ is a slowly varying function of time $t$, and $\h(t) \psi_n(t) = E_n(t) \psi_n(t)$, $(n=i,f)$, then the probability $\Gamma$ of the transition $\psi_i \to \psi_f$ is given by (atomic units $\hbar = m_e = |e| = 1$ are used throughout)
\begin{equation}\label{DykhneApp}
\Gamma = \exp[ -2 \Im(S)], \quad S = \int_{t_1}^{t_0} \Big[E_f(t) -E_i(t)\Big] dt,
\end{equation}
where $t_1$ is any point on the real axis of $t$, and $t_0$ is the transition point, i.e.,  a complex root of the equation
\begin{equation}\label{EqT0}
E_i(t_0) = E_f(t_0),
\end{equation}
which lies in the upper half-plane. If there are several roots, we must choose one that is the closest to the real axis of $t$.  Moreover, there are no assumptions regarding the form of the Hamiltonian. 

Now we shall apply the Dykhne approach to the problem of ionization of a single electron under the influence of a linearly polarized laser field with the frequency $\omega$ and the strength $\mathbf{F}$. The initial and final energies for such a process are given by 
\begin{subequations}
\begin{equation}
E_i (t) = -I_p,
\end{equation}
\begin{equation}
E_f (t) = \frac 12 [\K + \A(t)]^2,
\end{equation}
\end{subequations}
where $I_p$ is the ionization potential, $\K$ is the canonical momentum (measured on the detector), and $\A(t) = - \frac {\mathbf{F}}{\omega}\sin(\omega t)$.

According to Eq. (\ref{DykhneApp}), the probability of one-electron ionization $\Gamma$ can be written as 
\begin{eqnarray}\label{SPAEq1}
\Gamma &=& \exp\left[ -2\Im S \left( \kk, \k, I_p\right)\right], \\
S \left( \kk, \k, I_p\right) &=& \int_{t_1}^{t_0}\left( \frac 12 [\kk + A(t)]^2 + \frac 12 \k^2 + I_p \right) dt, \nonumber
\end{eqnarray}
where $S$ is the action. Equation (\ref{EqT0}) can be rewritten in terms of $S$ as
\begin{equation}\label{SPAEq2}
\frac{\partial}{\partial t_0} S \left( \kk, \k, I_p\right) = 0.
\end{equation}
Note that the analogy between the saddle point S-matrix calculations \cite{Ivanov_2005}, where transitions are calculated using stationary points of the action, and the Dykhne approach can be seen from Eqs. (\ref{SPAEq1}) and (\ref{SPAEq2}). The transition point is given by
\begin{equation}\label{ExSolutT0}
\omega t_0 =   \mathrm{Arcsin}\left[ \gamma\left( \frac{ \kk}{\sqrt{2I_p}} + i\sqrt{1 +\frac{\k^2}{2I_p}} \right)\right],
\end{equation}
where $\gamma$ is the Keldysh parameter
$$
\gamma = \frac{\omega}F\sqrt{2I_p} = \sqrt{\frac{I_p}{2U_p}}
$$
and $U_p = \left( \frac F{2\omega}\right)^2$ is the ponderomotive potential. In order to extract the imaginary and real parts of this solution, the following equation \cite{Abramowitz_1972} can be used 
\begin{equation}\label{AbramArcsin}
 \mathrm{Arcsin}\, (x+iy) = 2K\pi + \arcsin\beta + i\ln\left[\alpha + \sqrt{\alpha^2 -1}\right],
\end{equation}
where $K$ is an integer and
\begin{eqnarray}
\left\{ \alpha\atop \beta \right\} = \frac 12 \sqrt{(x+1)^2 + y^2} \pm \frac 12\sqrt{(x-1)^2 + y^2}.    \nonumber
\end{eqnarray}
Using Eq. (\ref{AbramArcsin}) in Eq. (\ref{SPAEq1}), we obtain
\begin{equation}\label{GeneralRate}
\Gamma (\gamma, \kk, \k)  \propto \exp\left[- \frac{2I_p}{\omega}f(\gamma, \kk, \k) \right],
\end{equation}
where
\begin{eqnarray}
f(\gamma, \kk, \k)  &=& \left(1+\frac 1{2\gamma^2}+ \frac{k^2}{2I_p} \right)\arccosh\alpha - \sqrt{\alpha^2 -1}\left( \frac{\beta}{\gamma}\sqrt{\frac 2{I_p}}\kk  +
\frac{\alpha\left[1-2\beta^2\right]}{2\gamma^2} \right),  \nonumber\\
\left\{ \alpha\atop \beta \right\} &=& \frac{\gamma}2 \left( 
\sqrt{\frac{k^2}{2I_p} + \frac 2{\gamma}\frac{\kk}{\sqrt{2I_p}} 
+ \frac 1{\gamma^2} +1 }   \pm
\sqrt{\frac{k^2}{2I_p} - \frac 2{\gamma}\frac{\kk}{\sqrt{2I_p}} 
+ \frac 1{\gamma^2} +1 }
\right), \nonumber\\
k^2 &=& \kk^2 + \k^2. \nonumber
\end{eqnarray}
Note that $\alpha > 1$. It must be stressed here that {\it no assumptions on the momentum of the electron have been made.} However, Eq. (\ref{GeneralRate}) has an exponential accuracy because the influence of the Coulomb field of a nucleus cannot be accounted for by the strong field approximation. The correct exponential prefactor has been obtained within the Perelomov-Popov-Terent'ev (PPT) approach \cite{Perelomov_1966, Perelomov_1967_A, Perelomov_1967_B, Perelomov_1968}.

%%% Reply to Comment 2 %%%%%%%%%%%
Similarly to the Yudin-Ivanov formula, Eq.  (\ref{GeneralRate}) is valid if the strength of the laser field $F$ depends on time, $F\to E_0 g(t)$, where the envelope  $g(t)$ of the pulse is assumed to be nearly constant during one-half of a laser cycle.
%%%%%%%%%%%%%%%%%%%%%%%%%

Equation (\ref{GeneralRate}) is the central result of this paper.  In the following, this equation is applied to  some special cases in order to establish connections with previous results. 

In the case of zero final momentum ($k=0$), we have that $\alpha = \sqrt{1+ \gamma^2}$ and $\beta = 0$. 
In this limit we recover the original Keldysh formula \cite{Keldysh_1965}
$$
f(\gamma,0,0) = \left( 1 + \frac 1{2\gamma^2} \right) \arcsinh \gamma -
\frac {\sqrt{1+\gamma^2}}{2\gamma}.
$$

In the tunneling limit ($\gamma\ll 1$) the following formulas can be obtained. Expanding the function $f(\gamma, \kk, \k) $ in a Taylor series up to third order with respect to $\gamma$ and setting $\k=0$, we obtain
\begin{equation}\label{CorkumTunnelEq}
\Gamma (\gamma, \kk, 0) \approx  \Gamma (\gamma, 0,0)
\exp\left[ -\frac{ \kk^2}{3\omega}\gamma^3 \right]. 
\end{equation}
Eq. (\ref{CorkumTunnelEq}) has been derived by a classical approach in Ref. \cite{Corkum_1989} (see also Ref. \cite{Delone_1991}). Discussions regarding the physical origin of Eq. (\ref{CorkumTunnelEq}) are presented in Ref. \cite{Ivanov_2005}. Performing the same expansion and setting $\kk=0$, we obtain
\begin{equation}\label{DeloneTunnelEq}
\Gamma (\gamma, 0,\k) \propto \exp\left[ -\frac{2\left(\k^2 + 2I_p\right)^{3/2}}{3F}\right].
\end{equation} 
This equation has been derived in Ref. \cite{Delone_1991}. 
For small values of $\k$, Eq. (\ref{DeloneTunnelEq}) can be approximated by
\begin{equation}\label{IvanovTunnelEq}
\Gamma (\gamma, 0,\k) \approx  \Gamma (\gamma, 0,0)
\exp\left[ -\frac{\sqrt{2I_p}\k^2}F \right].
\end{equation}

Let us fix $\k=0$ and continue working in the tunneling regime. For the case of high kinetic energy, such as $\kk^2/2 > 2U_p$ and $\sqrt{\kk^2/(4U_p)}-1 \gg \gamma$, we obtain 
$$
\alpha \approx \sqrt{ \frac{\kk^2}{4U_p}}, \qquad \beta \approx 1,
$$
and the ionization rate $\Gamma$ is given by 
\begin{equation}\label{KrainovEq7}
\Gamma \propto \exp\left\{ -\frac{2U_p}{\omega}\left[ \left( \frac{\kk^2}{2U_p} + 1\right) \arccosh\sqrt{\frac{\kk^2}{4U_p}} - 
3\sqrt{\frac{\kk^2}{4U_p}\left(\frac{\kk^2}{4U_p} -1 \right)}
\right] \right\}.
\end{equation}
Eq. (\ref{KrainovEq7}) has been obtained in Ref. \cite{Krainov_2003}.

Calculating the asymptotic expansion of the function $f(\gamma, \kk, 0) $ for $\kk^2/2 \gg 2U_p$, we obtain
\begin{equation}\label{KrainovTunnelEq}
\Gamma \propto \left( \frac{F^2 \exp(3)}{4\omega^2 \kk^2} \right)^{\frac{\kk^2}{2\omega}} \qquad (\kk^2/2 \gg 2U_p, \,\k=0).
\end{equation}
Eq. (\ref{KrainovTunnelEq}) has been obtained for tunneling ionization in Ref. \cite{Krainov_2003}. Here, we have proved that {\it Eq. (\ref{KrainovTunnelEq}) is valid for arbitrary  $\gamma$.} A similar formula can be derived for $\k\neq 0$
\begin{equation}\label{PPerpAsympt}
\Gamma \propto \left( \frac{F^2 \exp(1)}{4\omega^2\k^2} \right)^{\frac{\k^2}{2\omega}} \qquad (\k^2/2 \gg 2U_p, \, \kk=0),
\end{equation}
which is also valid for arbitrary values of the Keldysh parameter $\gamma$.

Consider the asymptotic expansion of Eq. (\ref{GeneralRate}) for large values of $I_p$ ($I_p \gg k^2/2$). In this case $\alpha$ and $\beta$ can be approximated by
$$
\alpha \approx \sqrt{1+\gamma^2}, \qquad 
\beta \approx \frac {\gamma}{\sqrt{1+\gamma^2}} \frac{\kk}{\sqrt{2I_p}}.
$$
Using these equations, we obtain
\begin{equation}\label{PPTEq}
f(\gamma, \kk, \k) \approx f(\gamma, 0,0) + 
\frac{k^2}{2I_p}\arcsinh\gamma - 
\frac{\gamma}{\sqrt{1+\gamma^2}}\frac{\kk^2}{2I_p}.
\end{equation}
Eq. (\ref{PPTEq}) has been reached within the PPT approach \cite{Perelomov_1966, Perelomov_1967_A, Perelomov_1967_B, Perelomov_1968} (see also \cite{Popov_2004}). 

As mentioned above, Goreslavskii {\it et al.} \cite{Goreslavskii_2005} have derived an expression for the spectral-angular distribution of single-electron ionization without any assumptions on the momentum of the electron. However, they have summed over saddlepoints, i.e., the contribution from previous laser cycles has been taken into account. On the contrary, we have not performed any summation because we are interested in the most recent contribution to ionization. Therefore, our result does account for the phase dependence of the ionization rate, unlike that of
Ref. \cite{Goreslavskii_2005}.

To make the phase dependance explicit in Eq. (\ref{GeneralRate}), we apply the  substitution  
\begin{equation}\label{PhaseDep}
\kk \to \kk - A(t).
\end{equation}
The analytical expression for the ionization rate as a function of a laser phase when $k = 0$ has been achieved by Yuding and Ivanov \cite{Yudin_2001}. Thus, Eq.~(\ref{GeneralRate}) is seen to be a generalization of the Yudin-Ivanov formula.

%%%%% Reply to Comment 1 %%%%%
Note that generally speaking, there is no unique and consistent way of defining the instantaneous ionization rates within quantum mechanics, and such a definition is a topic of an ongoing discussion (see, e.g., Refs. \cite{Smirnova2008a, Saenz_2007} and references therein). However,
the instantaneous ionization rates are indeed rigorously defined within the quasiclassical approximation (the Yudin-Ivanov formula), and we have employed this approach in the current paper. Alternatively, one can approximate the instantaneous ionization rates by the static ionization rates at each point in time using the instantaneous value of the laser field.
%%%%%%%%%%%%%%%%%%%%%

\begin{figure}
\begin{center}
\includegraphics[scale=0.4]{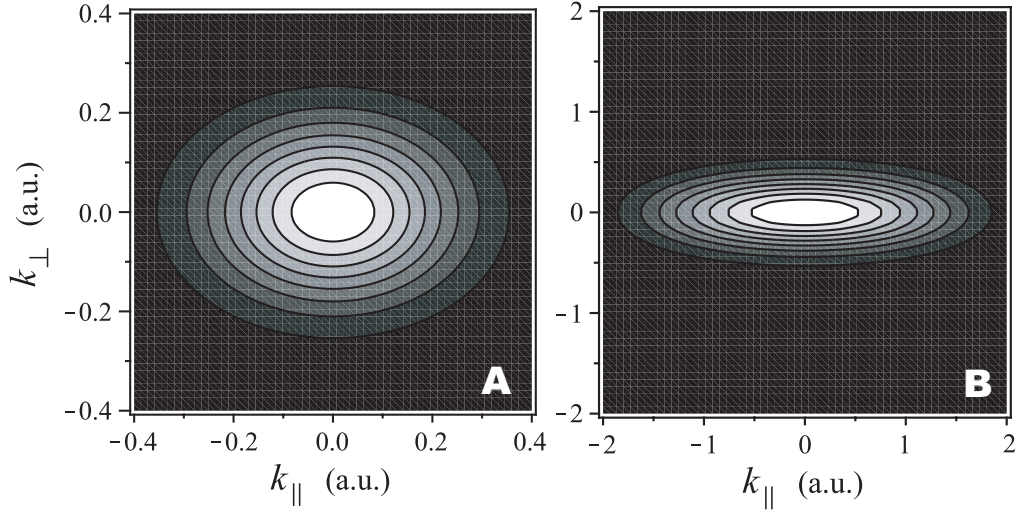}
\end{center}
\caption{The single-electron spectrum of a hydrogen atom at $800$ nm; plot A: at $1\times 10^{13}$ W/cm$^2$ ($\gamma = 3.376$); plot B at $6\times 10^{14}$ W/cm$^2$ ($\gamma=0.4357$).}\label{Fig1}
\end{figure}

\begin{figure}
\begin{center}
\includegraphics[scale=0.4]{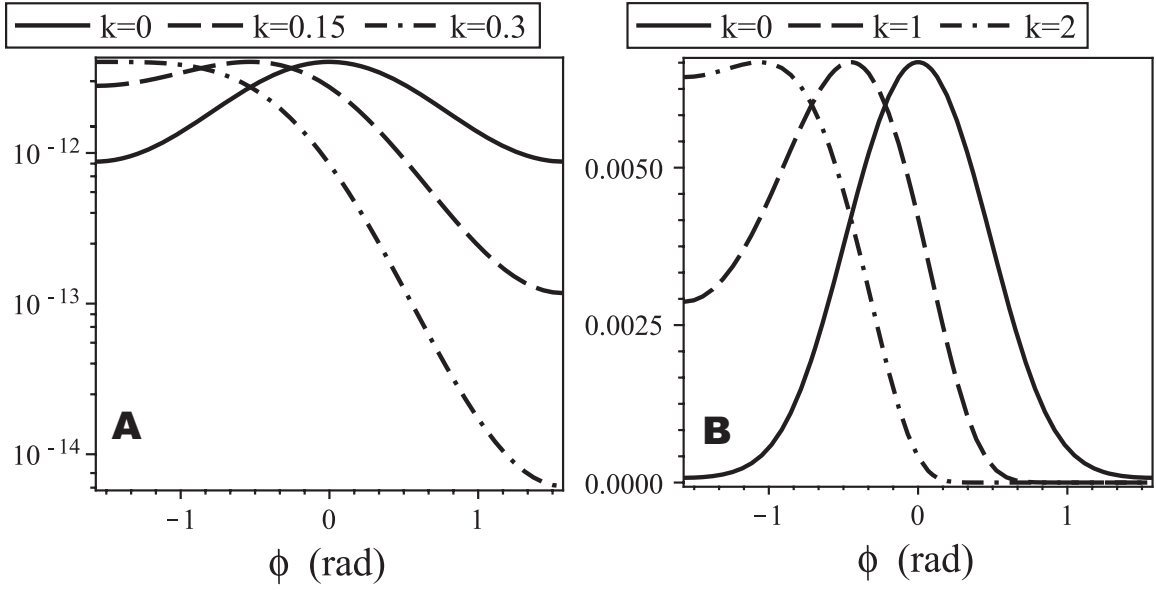}
\end{center}
\caption{The plot of $\Gamma(\gamma, k+\frac F{\omega}\sin\phi, 0)$ for a hydrogen atom at $800$ nm; plot A: at $1\times 10^{13}$ W/cm$^2$ ($\gamma = 3.376$); plot B at $6\times 10^{14}$ W/cm$^2$ ($\gamma=0.4357$).}\label{Fig2}
\end{figure}

\begin{figure}
\begin{center}
\includegraphics[scale=0.4]{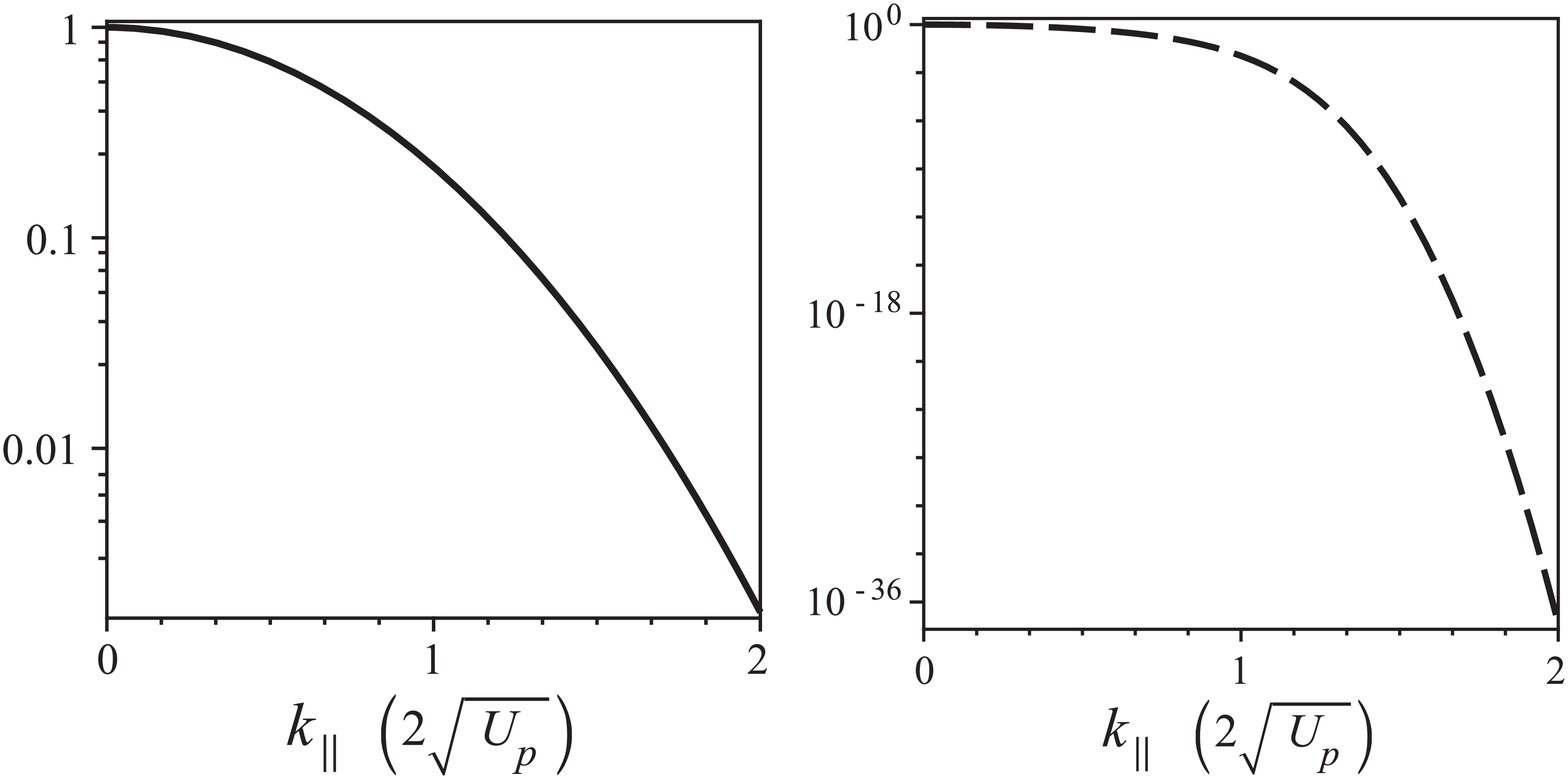}
\end{center}
\caption{The plot of $\Gamma(\gamma, \kk, 0)/\Gamma(\gamma,0,0)$ for a hydrogen atom at $800$ nm; the solid line: $1\times 10^{13}$ W/cm$^2$ ($\gamma = 3.376$); the dashed line: $6\times 10^{14}$ W/cm$^2$ ($\gamma=0.4357$).}\label{Fig3}
\end{figure}

Lastly, we illustrate Eq. (\ref{GeneralRate}) for the case of a hydrogen atom. The single-electron ionization spectra in the multiphoton regime ($\gamma \gg 1$) and in the tunneling regime ($\gamma \ll 1$) are plotted in Fig. \ref{Fig1}.A and Fig. \ref{Fig1}.B respectively. One concludes that the smaller $\gamma$, the more elongated the single-electron spectrum. 
%%%%%% Reply to Comment 3 %%%%%
We can notice that the maxima of both the spectra are at the origin. Nevertheless, a dip at the origin has been observed experimentally \cite{Rudenko2004a, Alnaser2006} in the parallel-momentum
distribution for the nobel gases within the tunneling regime, and afterwards it has been investigated theoretically in Ref. \cite{Guo2008} and references therein. However, such a phenomenon is beyond Eq. (\ref{GeneralRate}).
%%%%%%%%%%%%%%%%%%%%%%
The phase dependence of ionization for different initial momenta, recovered by means of Eq. (\ref{PhaseDep}), is illustrated in Fig. \ref{Fig2} for selected positive momenta. The curves for negative momenta are mirror reflections (through the axis $\phi = 0$) of the corresponding positive curves. Figure \ref{Fig3} shows that the cutoff of the single-electron spectrum in the tunneling regime (the dashed line) corresponds exactly to the kinetic energy $2U_p$, which is the maximum kinetic energy of a classical electron oscillating under the influence of a linearly polarized laser field.

In summary, we have derived a novel expression for strong field ionization including both the dependence on momenta and instantaneous laser phase.   Previous results \cite{Yudin_2001, Keldysh_1965, Krainov_2003, Delone_1991} regarding strong field ionization can be recovered as special cases of the general formula.  The present result concerns only the exponential dependence of the ionization process.  However, supplementing our approach with the pre-exponential factor taken from PPT \cite{Perelomov_1966, Perelomov_1967_A, Perelomov_1967_B, Perelomov_1968, Popov_2004}, the present result is the most general formula obtained within the quasi-lassical approximation.

The author is grateful to M. Spanner, G. L. Yudin, and M. Yu. Ivanov for many valuable discussions and remarks, and for outstanding comments on the manuscript.

\bibliography{OneElectron}

\begin{thebibliography}{25}
\expandafter\ifx\csname natexlab\endcsname\relax\def\natexlab#1{#1}\fi
\expandafter\ifx\csname bibnamefont\endcsname\relax
  \def\bibnamefont#1{#1}\fi
\expandafter\ifx\csname bibfnamefont\endcsname\relax
  \def\bibfnamefont#1{#1}\fi
\expandafter\ifx\csname citenamefont\endcsname\relax
  \def\citenamefont#1{#1}\fi
\expandafter\ifx\csname url\endcsname\relax
  \def\url#1{\texttt{#1}}\fi
\expandafter\ifx\csname urlprefix\endcsname\relax\def\urlprefix{URL }\fi
\providecommand{\bibinfo}[2]{#2}
\providecommand{\eprint}[2][]{\url{#2}}

\bibitem[{\citenamefont{Keldysh}(1965)}]{Keldysh_1965}
\bibinfo{author}{\bibfnamefont{L.~V.} \bibnamefont{Keldysh}},
  \bibinfo{journal}{Sov. Phys. JETP} \textbf{\bibinfo{volume}{20}},
  \bibinfo{pages}{1307} (\bibinfo{year}{1965}).

\bibitem[{\citenamefont{Yudin and Ivanov}(2001)}]{Yudin_2001}
\bibinfo{author}{\bibfnamefont{G.~L.} \bibnamefont{Yudin}} \bibnamefont{and}
  \bibinfo{author}{\bibfnamefont{M.~Y.} \bibnamefont{Ivanov}},
  \bibinfo{journal}{Phys. Rev. A} \textbf{\bibinfo{volume}{64}},
  \bibinfo{pages}{013409} (\bibinfo{year}{2001}).

\bibitem[{\citenamefont{Uiberacker and et~al.}(2007)}]{Uiberacker_2007}
\bibinfo{author}{\bibfnamefont{M.}~\bibnamefont{Uiberacker}} \bibnamefont{and}
  \bibinfo{author}{\bibnamefont{et~al.}}, \bibinfo{journal}{Nature}
  \textbf{\bibinfo{volume}{446}}, \bibinfo{pages}{627} (\bibinfo{year}{2007}).

\bibitem[{\citenamefont{Kienberger et~al.}(2007)\citenamefont{Kienberger,
  Uiberacker, Kling, and Krausz}}]{Kienberger_2007}
\bibinfo{author}{\bibfnamefont{R.}~\bibnamefont{Kienberger}},
  \bibinfo{author}{\bibfnamefont{M.}~\bibnamefont{Uiberacker}},
  \bibinfo{author}{\bibfnamefont{M.~F.} \bibnamefont{Kling}}, \bibnamefont{and}
  \bibinfo{author}{\bibfnamefont{F.}~\bibnamefont{Krausz}},
  \bibinfo{journal}{J. Mod. Opt.} \textbf{\bibinfo{volume}{54}},
  \bibinfo{pages}{1985} (\bibinfo{year}{2007}).

\bibitem[{\citenamefont{Krainov}(2003)}]{Krainov_2003}
\bibinfo{author}{\bibfnamefont{V.~P.} \bibnamefont{Krainov}},
  \bibinfo{journal}{J. Phys. B} \textbf{\bibinfo{volume}{36}},
  \bibinfo{pages}{L169} (\bibinfo{year}{2003}).

\bibitem[{\citenamefont{Goreslavskii et~al.}(2005)\citenamefont{Goreslavskii,
  Popruzhenko, Shvetsov-Shilovskii, and Shcherbachev}}]{Goreslavskii_2005}
\bibinfo{author}{\bibfnamefont{S.~P.} \bibnamefont{Goreslavskii}},
  \bibinfo{author}{\bibfnamefont{S.~V.} \bibnamefont{Popruzhenko}},
  \bibinfo{author}{\bibfnamefont{N.~I.} \bibnamefont{Shvetsov-Shilovskii}},
  \bibnamefont{and} \bibinfo{author}{\bibfnamefont{O.~V.}
  \bibnamefont{Shcherbachev}}, \bibinfo{journal}{JETP}
  \textbf{\bibinfo{volume}{100}}, \bibinfo{pages}{22} (\bibinfo{year}{2005}).

\bibitem[{\citenamefont{Popov}(2004)}]{Popov_2004}
\bibinfo{author}{\bibfnamefont{V.~S.} \bibnamefont{Popov}},
  \bibinfo{journal}{Physics -- Uspekhi} \textbf{\bibinfo{volume}{47}},
  \bibinfo{pages}{855} (\bibinfo{year}{2004}).

\bibitem[{\citenamefont{Delone and Krainov}(1985)}]{Delone_1985}
\bibinfo{author}{\bibfnamefont{N.~B.} \bibnamefont{Delone}} \bibnamefont{and}
  \bibinfo{author}{\bibfnamefont{V.~P.} \bibnamefont{Krainov}},
  \emph{\bibinfo{title}{Atoms in strong light fields}}
  (\bibinfo{publisher}{Berlin: Springer-Verlag}, \bibinfo{year}{1985}).

\bibitem[{\citenamefont{Dykhne}(1962)}]{Dykhne_1962}
\bibinfo{author}{\bibfnamefont{A.~M.} \bibnamefont{Dykhne}},
  \bibinfo{journal}{Sov. Phys. JETP} \textbf{\bibinfo{volume}{14}},
  \bibinfo{pages}{941} (\bibinfo{year}{1962}).

\bibitem[{\citenamefont{Davis and Pechukas}(1976)}]{Davis_1976}
\bibinfo{author}{\bibfnamefont{J.~P.} \bibnamefont{Davis}} \bibnamefont{and}
  \bibinfo{author}{\bibfnamefont{P.}~\bibnamefont{Pechukas}},
  \bibinfo{journal}{J. Chem. Phys.} \textbf{\bibinfo{volume}{64}},
  \bibinfo{pages}{3129} (\bibinfo{year}{1976}).

\bibitem[{\citenamefont{Chaplik}(1964)}]{Chaplik_1964}
\bibinfo{author}{\bibfnamefont{A.~V.} \bibnamefont{Chaplik}},
  \bibinfo{journal}{Sov. Phys. JETP} \textbf{\bibinfo{volume}{18}},
  \bibinfo{pages}{1046} (\bibinfo{year}{1964}).

\bibitem[{\citenamefont{Landau and Lifshitz}(1977)}]{Landau_1977}
\bibinfo{author}{\bibfnamefont{L.~D.} \bibnamefont{Landau}} \bibnamefont{and}
  \bibinfo{author}{\bibfnamefont{E.~M.} \bibnamefont{Lifshitz}},
  \emph{\bibinfo{title}{Quantum mechanics: non-relativistic theory}}
  (\bibinfo{publisher}{Oxford; Toronto: Pergamon}, \bibinfo{year}{1977}).

\bibitem[{\citenamefont{Ivanov et~al.}(2005)\citenamefont{Ivanov, Spanner, and
  Smirnova}}]{Ivanov_2005}
\bibinfo{author}{\bibfnamefont{M.~Y.} \bibnamefont{Ivanov}},
  \bibinfo{author}{\bibfnamefont{M.}~\bibnamefont{Spanner}}, \bibnamefont{and}
  \bibinfo{author}{\bibfnamefont{O.}~\bibnamefont{Smirnova}},
  \bibinfo{journal}{J. Mod. Opt.} \textbf{\bibinfo{volume}{52}},
  \bibinfo{pages}{165} (\bibinfo{year}{2005}).

\bibitem[{\citenamefont{Abramowitz and Stegun}(1972)}]{Abramowitz_1972}
\bibinfo{author}{\bibfnamefont{M.}~\bibnamefont{Abramowitz}} \bibnamefont{and}
  \bibinfo{author}{\bibfnamefont{I.}~\bibnamefont{Stegun}},
  \emph{\bibinfo{title}{Handbook of Mathematical Functions}}
  (\bibinfo{publisher}{Dover Publications, Inc., New York},
  \bibinfo{year}{1972}).

\bibitem[{\citenamefont{Perelomov et~al.}(1966)\citenamefont{Perelomov, Popov,
  and Terent'ev}}]{Perelomov_1966}
\bibinfo{author}{\bibfnamefont{A.~M.} \bibnamefont{Perelomov}},
  \bibinfo{author}{\bibfnamefont{V.~S.} \bibnamefont{Popov}}, \bibnamefont{and}
  \bibinfo{author}{\bibfnamefont{M.~V.} \bibnamefont{Terent'ev}},
  \bibinfo{journal}{Sov. Phys. JETP} \textbf{\bibinfo{volume}{23}},
  \bibinfo{pages}{924} (\bibinfo{year}{1966}).

\bibitem[{\citenamefont{Perelomov et~al.}(1967)\citenamefont{Perelomov, Popov,
  and Terent'ev}}]{Perelomov_1967_A}
\bibinfo{author}{\bibfnamefont{A.~M.} \bibnamefont{Perelomov}},
  \bibinfo{author}{\bibfnamefont{V.~S.} \bibnamefont{Popov}}, \bibnamefont{and}
  \bibinfo{author}{\bibfnamefont{M.~V.} \bibnamefont{Terent'ev}},
  \bibinfo{journal}{Sov. Phys. JETP} \textbf{\bibinfo{volume}{24}},
  \bibinfo{pages}{207} (\bibinfo{year}{1967}).

\bibitem[{\citenamefont{Perelomov and Popov}(1967)}]{Perelomov_1967_B}
\bibinfo{author}{\bibfnamefont{A.~M.} \bibnamefont{Perelomov}}
  \bibnamefont{and} \bibinfo{author}{\bibfnamefont{V.~S.} \bibnamefont{Popov}},
  \bibinfo{journal}{Sov. Phys. JETP} \textbf{\bibinfo{volume}{25}},
  \bibinfo{pages}{336} (\bibinfo{year}{1967}).

\bibitem[{\citenamefont{Perelomov and Popov}(1968)}]{Perelomov_1968}
\bibinfo{author}{\bibfnamefont{A.~M.} \bibnamefont{Perelomov}}
  \bibnamefont{and} \bibinfo{author}{\bibfnamefont{V.~S.} \bibnamefont{Popov}},
  \bibinfo{journal}{Sov. Phys. JETP} \textbf{\bibinfo{volume}{26}},
  \bibinfo{pages}{222} (\bibinfo{year}{1968}).

\bibitem[{\citenamefont{Corkum et~al.}(1989)\citenamefont{Corkum, Burnett, and
  Brunel}}]{Corkum_1989}
\bibinfo{author}{\bibfnamefont{P.~B.} \bibnamefont{Corkum}},
  \bibinfo{author}{\bibfnamefont{N.~H.} \bibnamefont{Burnett}},
  \bibnamefont{and} \bibinfo{author}{\bibfnamefont{F.}~\bibnamefont{Brunel}},
  \bibinfo{journal}{Phys. Rev. Lett.} \textbf{\bibinfo{volume}{62}},
  \bibinfo{pages}{1259} (\bibinfo{year}{1989}).

\bibitem[{\citenamefont{Delone and Krainov}(1991)}]{Delone_1991}
\bibinfo{author}{\bibfnamefont{N.~B.} \bibnamefont{Delone}} \bibnamefont{and}
  \bibinfo{author}{\bibfnamefont{V.~P.} \bibnamefont{Krainov}},
  \bibinfo{journal}{J. Opt. Soc. Am. B} \textbf{\bibinfo{volume}{8}},
  \bibinfo{pages}{1207} (\bibinfo{year}{1991}).

\bibitem[{\citenamefont{Smirnova et~al.}(2006)\citenamefont{Smirnova, Spanner,
  and Ivanov}}]{Smirnova2008a}
\bibinfo{author}{\bibfnamefont{O.}~\bibnamefont{Smirnova}},
  \bibinfo{author}{\bibfnamefont{M.}~\bibnamefont{Spanner}}, \bibnamefont{and}
  \bibinfo{author}{\bibfnamefont{M.}~\bibnamefont{Ivanov}},
  \bibinfo{journal}{J. Phys. B} \textbf{\bibinfo{volume}{39}},
  \bibinfo{pages}{S307} (\bibinfo{year}{2006}).

\bibitem[{\citenamefont{Saenz and Awasthi}(2007)}]{Saenz_2007}
\bibinfo{author}{\bibfnamefont{A.}~\bibnamefont{Saenz}} \bibnamefont{and}
  \bibinfo{author}{\bibfnamefont{M.}~\bibnamefont{Awasthi}},
  \bibinfo{journal}{Phys. Rev. A} \textbf{\bibinfo{volume}{76}},
  \bibinfo{eid}{067401} (\bibinfo{year}{2007}).

\bibitem[{\citenamefont{Rudenko et~al.}(2004)\citenamefont{Rudenko, Zrost,
  Schr{\"o}ter, de~Jesus, Feuerstein, Moshammer, and Ullrich}}]{Rudenko2004a}
\bibinfo{author}{\bibfnamefont{A.}~\bibnamefont{Rudenko}},
  \bibinfo{author}{\bibfnamefont{K.}~\bibnamefont{Zrost}},
  \bibinfo{author}{\bibfnamefont{C.~D.} \bibnamefont{Schr{\"o}ter}},
  \bibinfo{author}{\bibfnamefont{V.~L.~B.} \bibnamefont{de~Jesus}},
  \bibinfo{author}{\bibfnamefont{B.}~\bibnamefont{Feuerstein}},
  \bibinfo{author}{\bibfnamefont{R.}~\bibnamefont{Moshammer}},
  \bibnamefont{and} \bibinfo{author}{\bibfnamefont{J.}~\bibnamefont{Ullrich}},
  \bibinfo{journal}{J. Phys. B} \textbf{\bibinfo{volume}{37}},
  \bibinfo{pages}{L407} (\bibinfo{year}{2004}).

\bibitem[{\citenamefont{Alnaser et~al.}(2006)\citenamefont{Alnaser, Maharjan,
  Wang, and Litvinyuk}}]{Alnaser2006}
\bibinfo{author}{\bibfnamefont{A.~S.} \bibnamefont{Alnaser}},
  \bibinfo{author}{\bibfnamefont{C.~M.} \bibnamefont{Maharjan}},
  \bibinfo{author}{\bibfnamefont{P.}~\bibnamefont{Wang}}, \bibnamefont{and}
  \bibinfo{author}{\bibfnamefont{I.~V.} \bibnamefont{Litvinyuk}},
  \bibinfo{journal}{J. Phys. B} \textbf{\bibinfo{volume}{39}},
  \bibinfo{pages}{L323} (\bibinfo{year}{2006}).

\bibitem[{\citenamefont{Guo et~al.}(2008)\citenamefont{Guo, Chen, Liu, and
  Gu}}]{Guo2008}
\bibinfo{author}{\bibfnamefont{L.}~\bibnamefont{Guo}},
  \bibinfo{author}{\bibfnamefont{J.}~\bibnamefont{Chen}},
  \bibinfo{author}{\bibfnamefont{J.}~\bibnamefont{Liu}}, \bibnamefont{and}
  \bibinfo{author}{\bibfnamefont{Y.~Q.} \bibnamefont{Gu}},
  \bibinfo{journal}{Phys. Rev. A} \textbf{\bibinfo{volume}{77}},
  \bibinfo{pages}{033413} (\bibinfo{year}{2008}).

\end{thebibliography}
\end{document}